\documentclass[aps,prd,preprint,nofootinbib]{revtex4}
\usepackage{amsfonts}
\parskip + 5pt
\usepackage{epsfig}
\usepackage{graphicx}

\begin{document}
\title{Testing Charmonium Production Mechanism via Polarized $J/\psi$
Pair Production at the LHC\\[7mm]}

\author{Cong-Feng Qiao$^{1,2}$, Li-Ping Sun$^{1}$, Peng Sun$^{1}$}
\affiliation{$^{1}$College of Physical Sciences, Graduate University
of Chinese Academy of Sciences \\ YuQuan Road 19A, Beijing 100049,
China} \affiliation{$^{2}$Theoretical Physics Center for Science
Facilities (TPCSF), CAS\\ YuQuan Road 19B, Beijing 100049, China}

\author{~\vspace{0.7cm}}

\begin{abstract}
At present the color-octet mechanism is still an important and
debatable part in the non-relativistic QCD(NRQCD). We find in this
work that the polarized double charmonium production at the LHC may
pose a stringent test on the charmonium production mechanism. Result
shows that the transverse momentum($p_T$) scaling behaviors of
double $J/\psi$ differential cross sections in color-singlet and
-octet production mechanisms deviate distinctively from each other
while $p_T$ is larger than 7 GeV. In color-octet mechanism, the two
$J/\psi$s in one pair are mostly transversely polarized when $p_T\gg
 2 m_c$, as expected from the fragmentation limit point of view.
In color-singlet mechanism, there is about one half of the
charmonium pairs with at least one $J/\psi$ being longitudinally
polarized at moderate transverse momentum. The energy dependence of
the polarized $J/\psi$ pair production is found to be weak, and this
process is found to be experimentally attainable in the early phase
of the LHC operation.
\\


\end{abstract}

\maketitle

\section{Introduction}
NRQCD\cite{nrqcd} has now become a basic theory to parameterize
non-perturbative contributions in heavy quarkonium production and
decays by color-singlet(CS) and color-octet(CO) matrix elements. It
defines the ``velocity scaling rules'' to order various matrix
elements by $v$, the relative velocity of heavy quarks in the rest
frame of heavy quarkonium. Although NRQCD has already achieved a lot
in the study of heavy quarkonium physics, there are still some
unsettled issues remaining, especially, in aspect of production
mechanism \cite{com}, for instance, the role of color-octet in
charmonium production.

The color-octet scheme is a central essence of NRQCD, which supplies
a way to partly understand the prompt $J/\psi$ and $\psi'$ surplus
production at the Fermilab Tevatron \cite{cdf}, and supplies a
consistent differential cross section spectrum, versus the
transverse momentum $p_T$, with experimental results \cite{Braaten}.
However, there are still something unclear in this scheme, which ask
for further effort. In electron-positron collision, the former
crises in $J/\psi$ exclusive
\cite{braaten02,b-factory-exclusive1,b-factory-exclusive2} and
inclusive production
\cite{b-factory-inclusive1,b-factory-inclusive2} are somehow going
to an end with the advent of the next-to-leading order(NLO) QCD
\cite{nloexclusive,nloinclusive} and relativistic correction results
\cite{nlorelative}. In charmonium photoproduction, the recent NLO
QCD calculation tells that to reproduce the full HERA experimental
results, the color-singlet contribution alone seems not enough
\cite{hera}. In charmonium hadroproduction, recent computations of
NLO QCD corrections \cite{nlohadropro} significantly enhance the
$J/\psi$ color-singlet yield in large transverse momentum region and
alter its polarization prediction in LO calculation. Hence, the NLO
$J/\psi$ hadroproduction result in CS scheme greatly minimize the
contribution through color-octet mechanism, which in some sense
agrees with the hypotheses given in Ref.\cite{hqqw}. For more
detailed discussion on these issues, readers can refer to recent
review articles, e.g., \cite{review1,review2,review3}.

To further investigate charmonium production mechanism keeps on
being an urgent and important task in the study of quarkonium
physics. The running of the LHC supplies a great opportunity to this
aim. With a luminosity of about $10^{32}\sim10^{34}cm^{-2}s^{-1}$
and center of mass energy $7 \sim 14$ TeV, LHC will produce copious
data of charmonium inclusive and exclusive production, which can in
principle answer the question: to what extent the color-octet
mechanism plays a role in $\psi$ production? As noted by Cho and
Wise that the direct $\psi$ production via color-octet mechanism
should be mostly transversely polarized at leading order in
$\alpha_s$, which may stand as a unique test to the quarkonium
production mechanism \cite{chowise}. In this work, we consider the
process of polarized $J/\psi$ pair production at LHC, and suggest to
test the charmonium production mechanism through it. The same
process but in unpolarized case was proposed and calculated by
Barger {\it et al}. \cite{barger} and Qiao \cite{qiao1} once for
Tevatron experiment.

The paper is organized as follows: In section II, we set up the
formalism for the computation of polarized $\psi$ hadroproduction,
in both color-singlet and -octet schemes. In section III, the
numerical results for the LHC first-year and later runs are
provided. The section IV is remained for a brief summary and
conclusions. For the convenience of readers, the relevant analytic
expressions are given in the Appendix.

\begin{figure}[b,m,u]
\centering
\includegraphics[width=6.5cm,height=5cm]{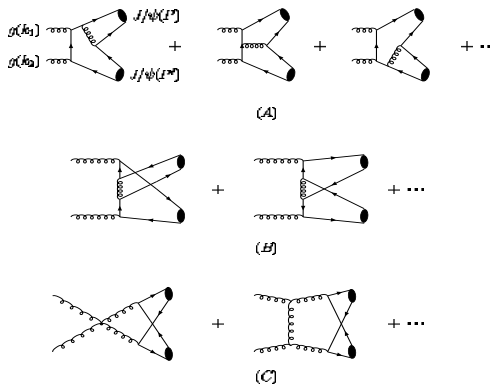}
\caption{\small Typical Feynman diagrams of $J/\psi$ pair production
in $p\,p$ collision at leading order in color-singlet scheme.}
\label{graph1}
\end{figure}

\section{Formalism}

\subsection{Color-Singlet Scheme}

In color-singlet model, the partonic subprocesses start at order
$\alpha_s^4$, which include $g + g \rightarrow J/\psi + J/\psi$ and
$q + \bar{q} \rightarrow J/\psi + J/\psi$. Intuitively, the latter,
the quark-antiquark annihilation process, is possibly negligible in
comparison with the former since the LHC is a proton-proton
collision machine. To confirm this point, the quark-antiquark
process is also evaluated in this work and the numerical result is
given in Table \ref{ratio1}. Hence, in the following analysis we
will mainly focus on the gluon-gluon process as shown in Figure
\ref{graph1}, and the analytic results of it are presented in the
Appendix.

For $J/\psi(^{3}S_{1})$ color-singlet hadronization, we employ the
following commonly used projection operator
\begin{eqnarray}
v(p_{\bar c})\,\overline{u}(p_c)& \longrightarrow& {1\over 2
\sqrt{2}} \not\! \epsilon^*_{J/\psi}\,(\not\!P+2 m_c)\, \times
\left( {1\over \sqrt{m_c}} \psi_{J/\psi}(0)\right) \otimes \left(
{{\bf 1}_c\over \sqrt{N_c}}\right)\, ,\label{eq:1}
\end{eqnarray}
where $\epsilon^{\mu}_{J/\psi}$ is the $J/\psi$ polarization vector
with $P\cdot \varepsilon=0$, ${\bf 1}_c$ stands for the unit color
matrix, and $N_c=3$. The nonperturbative parameter,
$\psi_{J/\psi}(0)$, is the Schr\"{o}dinger wave function of $c$ and
$\bar{c}$ system at the origin for $J/\psi$, and $|\psi_{J/\psi}(0)|
= \sqrt{ |R(0)|^2 /4\pi}$ with $R(0)$ being the radial wave function
at the origin. In our calculation, the non-relativistic relation
$m_{J/\psi} = 2 m_c$ is adopted, and hereafter, for simplicity
$m_{J/\psi}$ will be denoted by $m$.

The differential cross section for $J/\psi$ pair hadroproduction
reads as
\begin{eqnarray}
\label{eq:3} \frac{d\sigma}{d p_T} (p p \rightarrow 2J/\psi + X)  =
\sum_{a,b}\int dy_1 dy_2 f_{a/p}(x_a) f_{b/p}(x_b) 2 p_T x_a x_b
\frac{d\hat{\sigma}} {d{t}} (a + b \rightarrow 2J/\psi)\; ,
\end{eqnarray}
where $f_{a/p}$ and $f_{b/p}$ denote the parton densities; $y_1$,
$y_2$ are the rapidity of the two produced $J/\psi$s. 
The partonic scattering process, the gluon-gluon to polarized
$J/\psi$ pair $\frac{d\hat{\sigma}} {d{t}}$, can be calculated in
the standard method. To manipulate the trace and
matrix-element-square for those diagrams shown in Figure
\ref{graph1}, the computer algebra system MATHEMATICA is employed
with the help of the package FEYNCALC and FEYNARTS \cite{feyncalc}.
The lengthy expressions for different $J/\psi$ pair polarizations
are presented in the Appendix, for the convenience of independent
comparison and physical simulation use. Summing over all
polarizations, we find an agreement with the unpolarized cross
section given in \cite{qiao1}.

In experiment, the $J/\psi$ polarizations can be determined via
measuring the normalized angular distribution $I(\cos\theta)$ of
$\mu^+\mu^-$ pair in $J/\psi$ decays. The correlation of
distribution with polarization reads \cite{cdf2,spincorrelation}
\begin{eqnarray}
I(\cos\theta)=\frac{3}{2(\alpha+3)}(1+\alpha\; \cos^2\theta)\;
.\label{eq:10}
\end{eqnarray}
Here, $\theta$ is the angle between the $\mu^+$ direction in the
$J/\psi$ rest frame and the $J/\psi$ direction in pp center-of-mass
frame; $\alpha$ is a convenient measure of the polarization, which
is defined as $\alpha =(\sigma_T - 2 \sigma_L)/(\sigma_T + 2
\sigma_L)$ where $\sigma_T$ and $\sigma_L$ are transverse and
longitudinal components of the cross section, respectively. For
detailed derivation of the relation between polarization measure
$\alpha$ and decay distribution, see, e.g., the Appendix A of Ref.
\cite{spincorrelation}. From the definition, obviously, $\alpha = 1$
corresponds to the transverse polarization, and $\alpha= -1$ to the
longitudinal polarization.

\subsection{Color-Octet Scheme}

\begin{figure}[t,m,u]
\centering
\includegraphics[width=13cm,height=5cm]{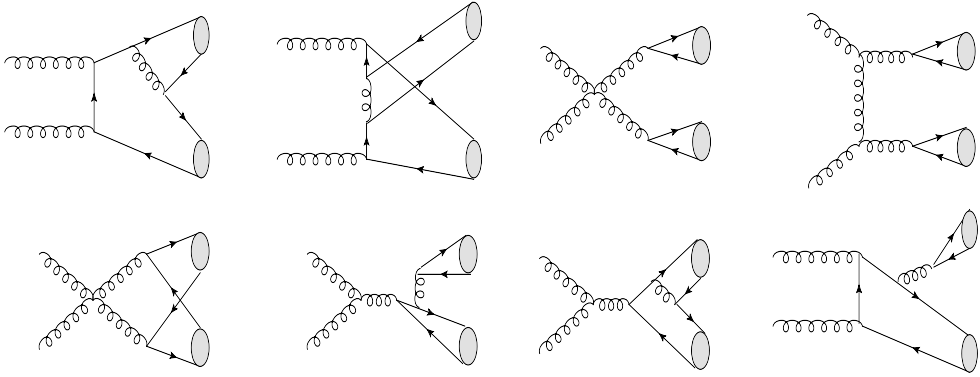}%
\caption{\small Typical Feynman diagrams of $J/\psi$ pair production
in $p\,p$ collision at leading order in color-octet scheme.}
\label{graph2}
\end{figure}
In NRQCD, the short-distant contribution can be expanded by $v$.
Dynamical gluons enter into Fock state, and combine with heavy quark
pair to form Color octet states. Therefore, wave functions of
$J/\psi$ is represented by
\begin{eqnarray}
|J/\psi\rangle=\textsl{O}(1)|c\bar{c}[^3S_1^{(1)}]\rangle +
\textsl{O}(v)|c\bar{c}[^3P_J^{(8)}]g\rangle +
\textsl{O}(v^2)|c\bar{c}[^3S_1^{(1,8)}]gg\rangle +
\textsl{O}(v^2)|c\bar{c}[^1S_0^{(8)}]g\rangle+...
\end{eqnarray}

In the CO scheme, at the leading order of $v$, three different CO
state components $|c\bar{c}[^3P_J^{(8)}]g\rangle$,
$|c\bar{c}[^3S_1^{(1,8)}]gg\rangle$ and
$|c\bar{c}[^1S_0^{(8)}]g\rangle$ may contribute to the double
$J/\psi$ production. In practice, for the sake of simplicity for
further analysis, we classify the intermediate Fock states into two
clusters: one contains $|c\bar{c}[^3S_1^{(1,8)}]gg\rangle$, the
other includes $|c\bar{c}[^1S_0^{(8)}]g\rangle$ and
$|c\bar{c}[^3P_J^{(8)}]g\rangle$. In this sense, the double $J/\psi$
Fock states may have following different combinations:
\begin{eqnarray}
|J/\psi\rangle|J/\psi\rangle &=&
|c\bar{c}[^3S_1^{(1)}]\rangle|c\bar{c}[^3S_1^{(1)}]\rangle +
\underbrace{|c\bar{c}[^3S_1^{(1)}]
\rangle|c\bar{c}[^3S_1^{(8)}]gg\rangle}_{Part_1}
\nonumber\\
&+&\underbrace{|c\bar{c}[^3S_1^{(1)}]\rangle(|c\bar{c}[^3P_J^{(8)}]g
\rangle+|c\bar{c}[^1S_0^{(8)}]g\rangle)}_{Part_2}+
\underbrace{|c\bar{c}[^3S_1^{(8)}]gg
\rangle|c\bar{c}[^3S_1^{(8)}]gg\rangle}_{Part_3}\nonumber\\
&+&\underbrace{|c\bar{c}[^3S_1^{(8)}]gg\rangle(|c\bar{c}[^3P_J^{(8)}]g
\rangle+|c\bar{c}[^1S_0^{(8)}]g\rangle)}_{Part_4}\nonumber\\
&+&\underbrace{(|c\bar{c}[^3P_J^{(8)}]g\rangle+|c\bar{c}
[^1S_0^{(8)}]g\rangle)(|c\bar{c}[^3P_J^{(8)}]g
\rangle+|c\bar{c}[^1S_0^{(8)}]g\rangle)}_{Part_5}
\label{state}\end{eqnarray}
In (\ref{state}) the leading one
indicates the pure color-singlet state, which is discussed in the
preceding subsection. The CO states involved components are then
divided into five sectors. In small transverse momentum $p_T$
region, where the NRQCD factorization theorem does not work well,
the five CO state related processes would contribute less to the
double charmonium production than from the pure CS process due to a
suppression in matrix elements of $v^4$ or $v^8$. In large $p_T$
region, however, the CO contribution, especially from
$|c\bar{c}[^3S_1^{(8)}]gg\rangle$ may exceed what from the CS,
because the relative smallness of CO matrix elements may be
compensated by the enhancement of the large propagators in
corresponding processes, which is similar to the case of inclusive
$\psi$ production in large transverse momentum region at the
Fermilab Tevatron \cite{Braaten,pcho}. In infinite transverse
momentum limit, it is well-known that the process of gluon
fragmenting into $|c\bar{c}[^3S_1^{(8)}]gg\rangle$ will dominate
over others in $\psi$ hadroproduction. Hence, one may still expect
here that in the double $J/\psi$ exclusive hadroproduction in the
large transverse momentum region, the bi-gluon fragmentation would
be the most important production mechanism. However, in moderate
transverse momentum case as in LHC experiment, to minimize the
uncertainties induced by the fragmentation mechanism we calculate
the complete Feynman diagrams for CO processes instead of
fragmentation simplification as shown in Figure \ref{graph2}.

For processes involving the CO components, that is $part_1$ to
$part_5$ in (\ref{state}), those contribute double $J/\psi$
production via intermediate states $|c\bar{c}[^1S_0^{(8)}]g\rangle$
and $|c\bar{c}[^3P_J^{(8)}]g\rangle$ are obviously negligible in
comparison with the CS process as shown in Figure \ref{graph1}.
Moreover, the $part_4$ involved processes may contribute less than
what from $part_1$ processes by $v^4$, an order. For $Part_1$
contribution, our explicit numerical calculation shows that it is
smaller than what from the CS by about two orders of magnitude in
all $p_T$ region. In all, the meaningful CO contribution mainly
comes from the $part_3$ involved processes, and which are what we
are going to evaluate.

For the $J/\psi$ intermediate Fock state
$|c\bar{c}[^3S_1^{(8)}]gg\rangle$, the projection operator is
\begin{eqnarray}
v(p_{\bar c})\,\overline{u}(p_c)& \longrightarrow& {R_{8}(0)\over
\sqrt{16\pi m_{c}}}T^{a}_{\overline{c}c} \not\!
\epsilon^*_{J/\psi}\,(\not\!P+2 m_c)\, ,\label{eq:1}
\end{eqnarray}
\begin{eqnarray}
|R_8(0)|^2=\frac{\pi}{6}<{\cal O}^{J/\psi}_8({}^3S_1)>\; .
\end{eqnarray}
Except for the difference in CO and CS non-perturbative matrix
element projections, the perturbative calculations of Feynman
diagrams for both CS and CO are similar.

\section{Numerical results}

In numerical calculation, we enforce the LHC experimental condition,
the pseudorapidity cut $|\eta(J/\psi)| < 2.2$, on the produced
charmonium pair. The cross sections and transverse momentum spectra
are evaluated under three conditions: the first one with
center-of-mass energy $\sqrt{S}=7$ TeV, the second one with
center-of-mass energy $\sqrt{S}=10$ TeV, the third one with
center-of-mass energy $\sqrt{S}=14$ TeV, which correspond to the
experimental situations of initial and following runs for hadron
study at the LHC \cite{chengm}. The planned collisions happen at the
LHC with luminosity of $\sim 10^{32}cm^{-2}s^{-1}$ in first two
cases, and $\sim10^{33}cm^{-2}s^{-1}$ in the third situation. In one
year running, $10^{7}s$ effective time, the corresponding integrated
luminosities are about $1(fb^{-1})$ and $10(fb^{-1})$, respectively.
The input parameters take the values \cite{pcho}
\begin{eqnarray}
m_c =  1.5\; \rm{GeV},\; |R(0)|^2 = 0.8\;\rm{GeV}^3,\; <{\cal
O}^{J/\psi}_8({}^3S_1)> = 0.012\;\rm{GeV}^3\;.\label{eq:2}
\end{eqnarray}
The typical energy scale in partonic interaction is set to be at
$m_T =\sqrt{m^2 + p_T^2}$, and hence the strong coupling is running
with transverse momentum.

With the above formulas and inputs, one can readily obtain the
polarized $J/\psi$ pair production rate at the LHC. In our numerical
calculation, the parton distribution functions(PDFs) of
CTEQ5L\cite{cteq} is used, and both renormalization scale in strong
coupling $\alpha_s$ and factorization scale in PDFs are set to be
$m_T$. The numerical results for integrated cross section
$\sigma(p\; {p}\rightarrow J/\psi J/\psi)$ with different $p_T$
lower bounds are presented in Table I for the CS production scheme,
and Table II for the CO scheme, where the the branching fraction of
$ B(J/\psi \to \mu^+ \mu^-)= 0.0597$ is taken into account.
\begin{table}
\begin{center}
\caption{The integrated cross sections of $J/\psi$ pair production
in color-singlet model under various transverse momentum lower cuts
at the center-of-mass energy $\sqrt{S}=$ 7 TeV, 10 TeV and 14 TeV.
Here, $\bot\bot$ represents for the situation of both $J/\psi$s
being transversely polarized, $\|\|$ for both $J/\psi$s being
longitudinally polarized, $\|\bot$ for one $J/\psi$ being
longitudinally polarized and the other being transversely polarized.
The $tot_{gg}$ and $tot_{q\overline{q}}$ denote the
polarization-summed cross sections of gluon-gluon fusion and
quark-antiquark annihilation processes, respectively. }\tiny
\vspace{1mm}
\begin{tabular}{|c||c|c|c|c|c|c|c|c|c|c|c|c|c|c|c|}
\hline\hline &\multicolumn{5}{|c|}{7TeV}&\multicolumn{5}{|c|}{10TeV}
&\multicolumn{5}{|c|}{14TeV}\\
\hline\hline $\sigma\setminus p_{Tcut}$&~3 GeV~ & ~4 GeV~ & ~5 GeV~
& ~6 GeV~ & ~7 GeV~ &~3 GeV~ & ~4 GeV~ & ~5 GeV~ & ~6 GeV~ & ~7 GeV~
&~3 GeV~ & ~4 GeV~ & ~5 GeV~ & ~6 GeV~ &
~7 GeV~\\
\hline\hline $\bot\bot$ & 2.07pb & 0.61pb & 0.20pb & 0.070pb &
0.027pb & 2.84pb & 0.85pb & 0.29pb & 0.10pb & 0.039pb & 3.81pb &
1.16pb & 0.38pb
& 0.14pb & 0.055pb \\
\hline\hline $\|\|$ &0.79pb &0.29pb &0.11pb &0.040pb &0.016pb &
1.09pb & 0.41pb & 0.15pb & 0.040pb & 0.023pb & 1.45pb & 0.55pb &
0.20pb
& 0.079pb & 0.033pb\\
\hline\hline $\|\bot$ & 2.16pb & 0.55pb & 0.14pb & 0.040pb & 0.012pb
& 2.96pb & 0.76pb & 0.20pb & 0.056pb & 0.018pb & 3.96pb & 1.04pb &
0.27pb
& 0.079pb & 0.024pb \\
\hline\hline $tot_{gg}$ & 4.96pb & 1.43pb & 0.44pb & 0.15pb &
0.055pb & 6.81pb & 2.00pb & 0.62pb & 0.21pb & 0.079pb & 9.12pb &
2.72pb
& 0.85pb & 0.29pb & 0.11pb\\
\hline\hline $tot_{q\overline{q}}$ & 0.040pb & 0.013pb & 4.27fb &
1.57fb & 0.64fb & 0.047pb & 0.015pb & 5.12fb & 1.90fb & 0.77fb &
0.056pb & 0.017pb
& 6.12fb & 2.28fb & 0.93fb\\
\hline\hline
\end{tabular}
\label{ratio1}
\end{center}
\end{table}

\begin{figure}
\centering
\includegraphics[width=1.05\textwidth]{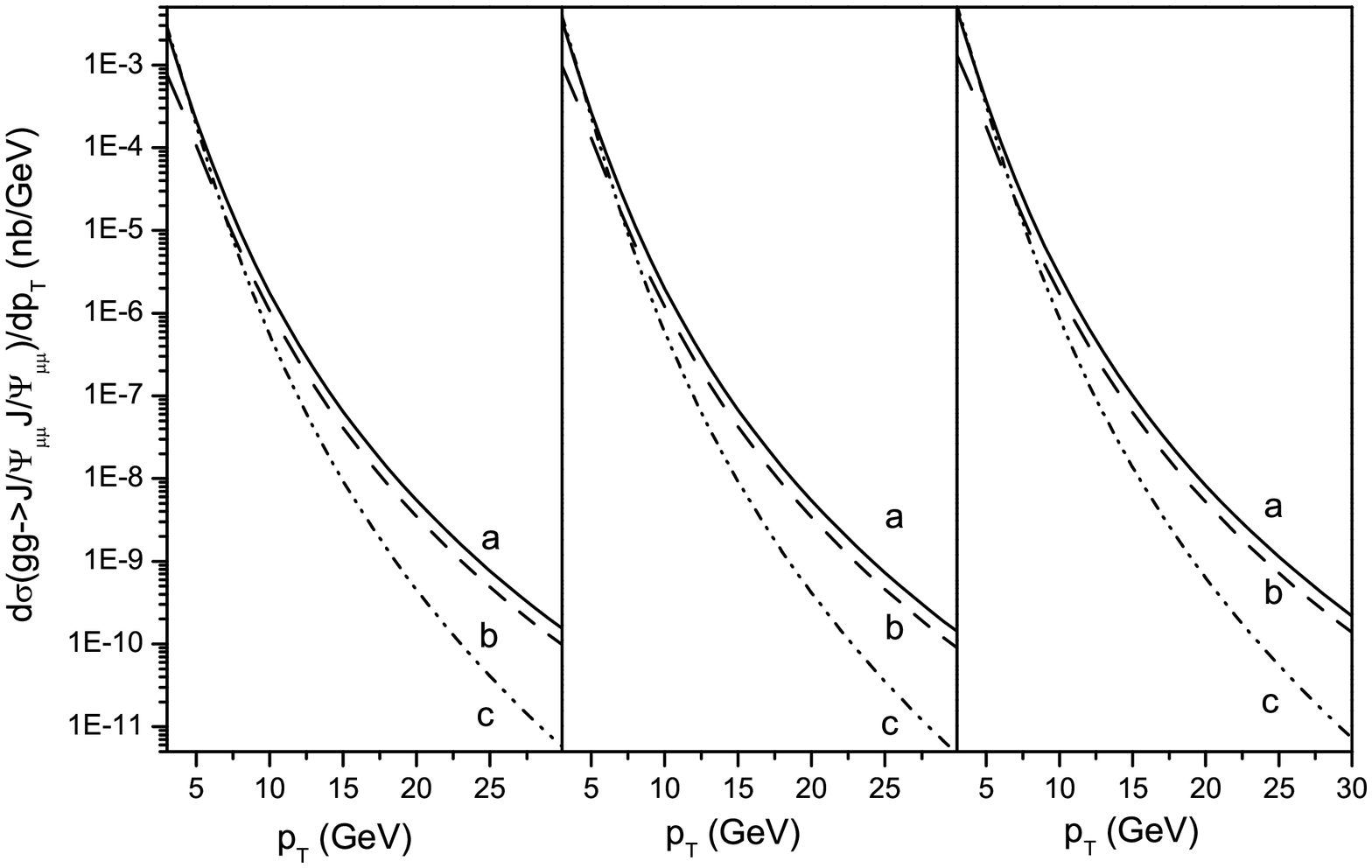}\hspace*{\fill}
\caption{\small The differential cross-section of $J/\psi$ pair
production versus $p_T$ at the LHC in color-singlet scheme. Lines
from top to bottom, i.e. a ,b and c, represent yields from
color-singlet in $\bot\bot$, $\|\|$, and $\|\bot$ cases,
respectively. The diagrams from left to right represent for cases of
$\sqrt{S}=7$ TeV, $\sqrt{S}=10$ TeV, and $\sqrt{S}=14$ TeV,
respectively.} \label{lpty1} \vspace{-0mm}
\end{figure}

\begin{figure}
\centering
\includegraphics[width=1.05\textwidth]{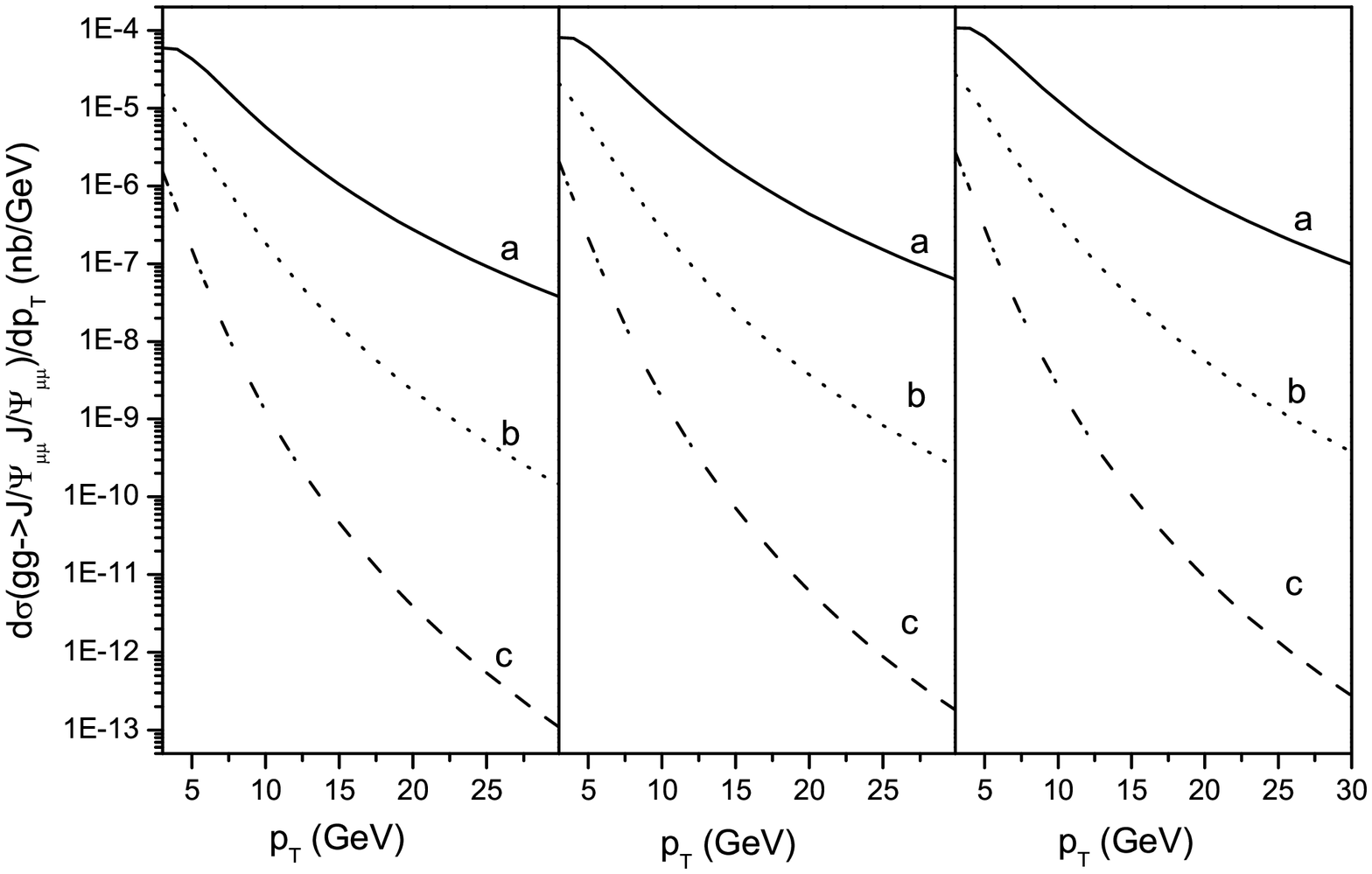}\hspace*{\fill}
\caption{\small The differential cross-section of $J/\psi$ pair
production versus $p_T$ at the LHC in color-octet scheme. Lines from
top to bottom, i.e. a ,b and c, represent yields from color-octet in
$\bot\bot$, $\|\bot$, and $\|\|$ cases, respectively.  The diagrams
from left to right represent for cases of $\sqrt{S}=7$ TeV,
$\sqrt{S}=10$ TeV, and $\sqrt{S}=14$ TeV, respectively.}
\label{lpty2} \vspace{-0mm}
\end{figure}

\begin{figure}
\centering
\includegraphics[width=1.05\textwidth]{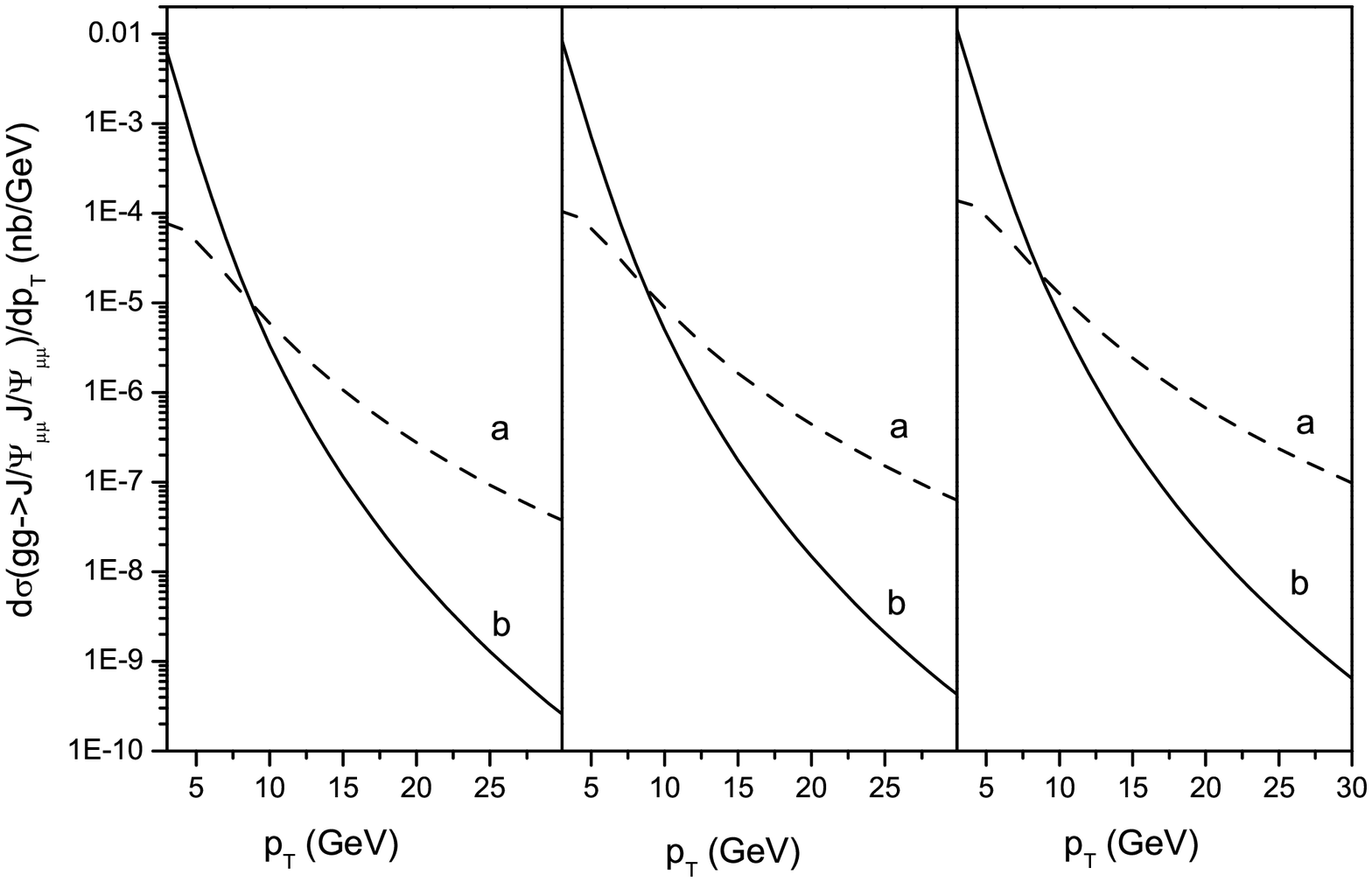}\hspace*{\fill}
\caption{\small The differential cross-section of $J/\psi$ pair
production versus $p_T$ at the LHC. Lines $a$ and $b$, represent
yields from color-octet and color-singlet in unpolarized case,
respectively.  The diagrams from left to right represent for cases
of $\sqrt{S}=7$ TeV, $\sqrt{S}=10$ TeV, and $\sqrt{S}=14$ TeV,
respectively.} \label{lpty3} \vspace{-0mm}
\end{figure}

\begin{table}
\begin{center}
\caption{The integrated cross sections of $J/\psi$ pair production
in color-octet model with various transverse momentum lower cuts at
the center-of-mass energy $\sqrt{S}=$ 7 TeV, 10 TeV and 14 TeV.
Here, $\bot\bot_{88}$ represents for the situation of both $J/\psi$s
being transversely polarized, $\|\|_{88}$ for both $J/\psi$s being
longitudinally polarized, $\|\bot_{88}$ for one $J/\psi$ being
longitudinally polarized and the other being transversely polarized,
and the $tot_{88}$ means the polarization-summed CO cross section in
gluon-gluon fusion. The $tot_{18}$ in last row represents for the
double $J/\psi$ yields from CS + CO production scheme for
reference.}\tiny \vspace{1mm}
\begin{tabular}{|c||c|c|c|c|c|c|c|c|c|c|c|c|c|c|c|}
\hline\hline &\multicolumn{5}{|c|}{7TeV}&\multicolumn{5}{|c|}{10TeV}
&\multicolumn{5}{|c|}{14TeV}\\
\hline\hline $\sigma\setminus p_{Tcut}$&~3 GeV~ & ~4 GeV~ & ~5 GeV~
& ~6 GeV~ & ~7 GeV~ &~3 GeV~ & ~4 GeV~ & ~5 GeV~ & ~6 GeV~ & ~7 GeV~
&~3 GeV~ & ~4 GeV~ & ~5 GeV~ & ~6 GeV~ &
~7 GeV~\\
\hline\hline $\bot\bot_{88}$ & 0.22pb & 0.16pb & 0.11pb & 0.075pb &
0.051pb & 0.31pb & 0.23pb & 0.16pb & 0.11pb & 0.076pb & 0.43pb &
0.32pb &
0.23pb & 0.16pb & 0.11pb\\
\hline\hline $\|\|_{88}$ & 1.33fb & 0.43fb & 0.14fb & 0.051fb &
0.020fb & 1.84fb & 0.60fb & 0.20fb & 0.073fb & 0.029fb & 2.46fb &
0.81fb & 0.28fb
& 0.10fb & 0.040fb\\
\hline\hline $\|\bot_{88}$ & 0.025pb & 0.013pb & 6.95fb & 3.62fb &
1.93fb & 0.035pb & 0.019pb & 9.95fb & 5.25fb & 2.84fb & 0.047pb &
0.026pb & 0.014pb
& 7.36fb & 4.02fb \\
\hline\hline $tot_{88}$ & 0.25pb & 0.18pb & 0.12pb & 0.080pb &
0.053pb & 0.35pb & 0.25pb & 0.17pb & 0.12pb & 0.079pb & 0.48pb &
0.35pb
& 0.24pb & 0.16pb & 0.11pb\\
\hline\hline $tot_{18}$ & 0.15pb & 0.047pb & 0.015pb & 5.14fb &
1.95fb & 0.21pb & 0.065pb & 0.021pb & 7.37fb & 2.83fb & 0.29pb &
0.089pb
& 0.029pb & 0.010pb & 3.97fb\\
\hline\hline
\end{tabular}
\label{ratio2}
\end{center}
\end{table}
The spectra of double-$J/\psi$ exclusive production as function of
transverse momentum $p_T$ are illustrated in Figures  \ref{lpty1} to
\ref{lpty3}. From these figures we see that the conventional CS
production scheme dominates over the CO one in relatively low-$p_T$
region, some $p_T < 7$ GeV.

\section{Summary and Conclusions}

We have in this work evaluated the polarized $J/\psi$ pair
hadroproduction rate, directly and exclusively, at the LHC in
color-singlet and -octet schemes. In color-octet scheme, Fock state
$|c\bar{c}[^3S_1^{(8)}]gg\rangle|c\bar{c}[^3S_1^{(8)}]gg\rangle$
involved process dominates over others, and it yields mainly the
transversely polarized $J/\psi$s. Other CO related processes, either
single or double CO processes, contribute significantly less than
those from CS and $part_1$ involved processes in whole transverse
momentum region, therefore can be neglected. Result shows that in
low transverse momentum region the double $J/\psi$ yields are
governed by the CS process, while the CO process may dominate as
$p_T > 7$ GeV. The indirect $J/\psi$ production from the $\chi_{cJ}$
and $\psi'$ feed down, according to the analysis of \cite{barger},
may contribute about a factor of 2 to the transverse distribution
spectrum of prompt $J/\psi$ production, while these inclusive yields
can be removed in experiment in principle and hence are not taken
into account in our analysis.

The calculation performed in this work is at the leading order of
strong coupling $\alpha_s$. Due to the relatively high interaction
scale in this process, higher order QCD corrections should be small.
The "smallness" here means that it may not be as big as NLO
corrections to some inclusive charmonium production processes
recently found. Nevertheless, this is just a naive speculation. For
exclusive process, higher order corrections, the virtual
corrections, do not change the relative polarization ratio, while is
different from NLO correction to the inclusive process where the
hard real correction may alter the final state polarization
significantly. Though higher order relativistic correction is
generally large in charmoniun system, while for $J/\psi$ pair
production it will be doubly suppressed. It is worthy of mentioning
that in order to keep the exclusivity of double $J/\psi$ production
and reduce the final state interaction effects, to enforce the
transverse momentum cut veto would be a necessary choice, however
which may also suppress some useful signals.

The integrated cross sections of double $J/\psi$ production with
different polarizations are found to be measurable theoretically
with a lower transverse momentum cut of 7 GeV in the first-year run
of LHC at colliding energy of 7 TeV and luminosity of 1$fb^{-1}$ or
so, where about one hundred of di-muons from $J/\psi$ decays will be
produced and the total yields from CS and CO are comparable. For
lower transverse momentum cut higher than $7$ GeV, the double
$J/\psi$ yield may mainly come from the color-octet scheme, under
the circumstance of employing the nowadays prevailing magnitude of
color-octet matrix element. At colliding energy of 7 TeV, our
calculation result shows that the data with at least one $J/\psi$ in
a pair being longitudinally polarized are nearly fifty percent of
the total yield with transverse momentum integration lower cut of 5
GeV of produced $J/\psi$. Different from the inclusive production,
where experimental data show that $J/\psi$s tend to be unpolarized
or even longitudinally polarized with the increase of transverse
momentum, in the exclusive pair production both color-singlet and
-octet schemes exhibit that the produced $J/\psi$ data should be
always dominated by the transversely polarized yields. The above
conclusions might be testified in future LHC experiment, and we
expect the experiment measurement on exclusive double $J/\psi$
production may tell us more on the charmonium production mechanism.

\vspace{.7cm} {\bf Acknowledgments} \vspace{.3cm}

This work was supported in part by the National Natural Science
Foundation of China(NSFC) under the grants 10935012, 10928510,
10821063 and 10775179, by the CAS Key Projects KJCX2-yw-N29 and
H92A0200S2, and by the Scientific Research Fund of GUCAS.

\newpage
\appendix{\bf\Large Appendix}

The polarized differential cross sections in color-singlet model.
Here, the symbol $\|\|$ represents for the situation of both
$J/\psi$ being longitudinally polarized; $\|\bot$ for one $J/\psi$
being longitudinally polarized and the other being transversely
polarized, and $\bot\bot$ for both $J/\psi$s being transversely
polarized. In these expressions, $s$, $t$, $u$ are normal Mandelstam
variables at the parton level.

\begin{eqnarray}
A^{\mu\nu\rho\sigma}_{ab}&=&\frac{64\alpha_s^2\pi
|R(0)|^2\delta_{ab}}
   {9m(m^2-u)^2(u+t-2m^2)^3} ((m^2-t) (27 m^6-(13 t+59 u) m^4+\nonumber\\
&&(-t^2+33 u t+31 u^2) m^2-18 t u^2) g^{\mu\sigma}
   g^{\nu\rho}-(p_4^\mu ((m^2-u) (3 m^2-2 t-u) p_4^\sigma m^2\nonumber\\
&&+(31 m^6-2 (15 t+34 u) m^4+(-2 t^2+70 u
   t+35 u^2) m^2-36 t u^2) p_3^\sigma)+2 p_2^\mu ((7 m^2\nonumber\\
&&+t-8 u) (m^2-u)
   p_3^\sigma m^2+(21 m^6-(13 t+47 u) m^4+(-t^2+33 u t+25 u^2) m^2\nonumber\\
&&-18 t u^2) p_4^\sigma))
   g^{\nu\rho}+(m^2-t) (m^2-u) (29 m^4+(2 t-13 u)
   m^2-18 t u) g^{\mu\rho} g^{\nu\sigma}\nonumber\\
&&-2 (m^2-t) (m^2-u) (15 m^4+(t-16 u)
   m^2+9 u (u-t)) g^{\mu\nu} g^{\rho\sigma}-(p_2^\mu (2 (28 m^6\nonumber\\
&&+(6 t-62 u) m^4+(-19 t^2+14 u t+33 u^2) m^2+18 t (t-u) u)
   p_2^\nu-(45 m^6+2 (4 t\nonumber\\
&&-67 u) m^4+(-38 t^2+68 u t+87 u^2) m^2+36 t (t-2 u) u)
   p_3^\nu)-(m^2-u) (11 m^4\nonumber\\
&&+(4 t+21 u) m^2-36 t u) p_4^\mu p_2^\nu)
g^{\rho\sigma}+g^{\mu\sigma}
   (2 p_2^\nu ((7 m^2+t-8 u) (m^2-u) p_4^\rho m^2\nonumber\\
&&+(21 m^6-(13 t+47 u) m^4+(-t^2+33 u t+25
   u^2) m^2-18 t u^2) p_3^\rho)-(45 m^6\nonumber\\
&&-14 (2 t+7 u) m^4+(-2 t^2+68 u t+51 u^2) m^2-36 t u^2)
   p_3^\nu (p_3^\rho+p_4^\rho))\nonumber\\
&&+g^{\nu\sigma} (2 p_2^\mu ((25 m^2-17 t-8
   u) (m^2-u) p_3^\rho m^2+(3 m^6+(23 t-11 u) m^4+(-19 t^2\nonumber\\
&&-21 u t+7 u^2) m^2+18 t^2 u)
   p_4^\rho)-(m^2-u) (11 m^4+(4 t+21 u) m^2-36 t u) p_4^\mu
   (p_3^\rho\nonumber\\
&&+p_4^\rho))-g^{\mu\rho} (2 p_2^\nu ((25 m^2-17 t-8 u) (m^2-u)
   p_4^\sigma m^2+(3 m^6+(23 t-11 u) m^4\nonumber\\
&&+(-19 t^2-21 u t+7 u^2) m^2+18 t^2 u)
   p_3^\sigma)+p_3^\nu ((5 m^6+(32 u-42 t) m^4+(38 t^2\nonumber\\
&&+2 u t-35 u^2) m^2+36 t u (u-t))
   p_3^\sigma-(m^2-u) (39 m^4-(38 t+37 u) m^2\nonumber\\
&&+36 t u) p_4^\sigma))+g^{\mu\nu} (p_4^\rho
   (2 (m^2-u) (16 m^2-17 t+u) p_4^\sigma m^2+(63 m^6-2 (14 t\nonumber\\
&&+67 u) m^4+(-38 t^2+104 u t+69 u^2) m^2+36 t (t-2
   u) u) p_3^\sigma)+(m^2-u) p_3^\rho (2 (16 m^2\nonumber\\
&&-17 t+u) p_3^\sigma m^2+(-7
   m^4+(4 t+39 u) m^2-36 t u) p_4^\sigma))-36
   (m^2-u) (p_2^\mu (2 (m^2\nonumber\\
&&-t)
   p_2^\nu (p_3^\rho p_4^\sigma-p_4^\rho p_3^\sigma)+p_3^\nu
   (p_4^\rho ((m^2-2 t+u) p_3^\sigma+2 (t-m^2) p_4^\sigma)-(m^2-2 t+u)
   p_3^\rho p_4^\sigma))\nonumber\\
&&+p_4^\mu (p_3^\nu
   (p_3^\rho+p_4^\rho) ((m^2-u) p_4^\sigma-(m^2-2 t+u)
   p_3^\sigma)+p_2^\nu ((m^2-u) p_4^\rho p_3^\sigma+p_3^\rho (2
   (m^2\nonumber\\
&&-t) p_3^\sigma+(u-m^2) p_4^\sigma)))))\label{eq:0}\end{eqnarray}
Here, $\mu$ and $\nu$ are polarization indices of outgoing
$J/\psi$s, $\rho$ and $\sigma$ are polarization indices of initial
gluons, $a$ and $b$ stand for colors.

\begin{eqnarray}
M^{\mu\nu\rho\sigma}_{ab}=A^{\mu\nu\rho\sigma}_{ab}
   +A_{ab}^{\nu\mu\rho\sigma}{(p_3\leftrightarrow
   p_4,u\leftrightarrow t)}
\end{eqnarray}

\begin{eqnarray}
\label{diff}  \frac{d\hat{\sigma}_{total}}{d {t}} &=& \frac{1}
{8192\pi
s^2}{\sum_{\lambda}}\sum_{ab}|(M\cdot\varepsilon(\lambda_{p_1})
\varepsilon(\lambda_{p_2})\varepsilon(\lambda_{p_3})
\varepsilon(\lambda_{p_4}))|^2\label{eq:5}
   \end{eqnarray}

\begin{eqnarray}
\label{diff}  \frac{d\hat{\sigma}_{\|\|}}{d {t}} &=& \frac{1}
{8192\pi
s^2}\sum_{\lambda}\sum_{ab}|(M\cdot\varepsilon
(\lambda_{p_1})\varepsilon(\lambda_{p_2})
\varepsilon(\lambda_{p_3})_{\|\|}\varepsilon
(\lambda_{p_4})_{\|\|})|^2\label{eq:6}
   \end{eqnarray}

\begin{eqnarray}
\label{diff}  \frac{d\hat{\sigma}_{\|\perp}}{d {t}} &=& \frac{1}
{8192\pi
s^2}\sum_{\lambda}\sum_{ab}(|(M\cdot\varepsilon
(\lambda_{p_1})\varepsilon(\lambda_{p_2})
 \varepsilon(\lambda_{p_3})_{\|\|}\varepsilon
 (\lambda_{p_4})_{\perp\perp})|^2\nonumber\\
&&+|(M\cdot\varepsilon(\lambda_{p_1})\varepsilon
(\lambda_{p_2})\varepsilon(\lambda_{p_3})
_{\perp\perp}\varepsilon(\lambda_{p_4})_{\|\|})|^2)\label{eq:6}
   \end{eqnarray}

\begin{eqnarray}
\label{diff}  \frac{d\hat{\sigma}_{\perp\perp}} {d {t}} &=&
\frac{1}{8192\pi
s^2}\sum_{\lambda}\sum_{ab}|(M\cdot\varepsilon
(\lambda_{p_1})\varepsilon(\lambda_{p_2})
\varepsilon(\lambda_{p_3})_{\perp\perp}\varepsilon
(\lambda_{p_4})_{\perp\perp})|^2\label{eq:7}
   \end{eqnarray}

\begin{eqnarray}
\sum_\lambda\varepsilon(\lambda_{p_2})^{\mu}*\varepsilon
(\lambda_{p_2})^{\nu}=\sum_\lambda
\varepsilon(\lambda_{p_1})^{\mu}*\varepsilon
(\lambda_{p_1})^{\nu}=-g^{\mu\nu}+\frac{p_1^{\mu}
p_2^{\nu}+p_2^{\mu}p_1^{\nu}}{p_1\cdot p_2}
\end{eqnarray}

\begin{eqnarray}
\sum_\lambda\varepsilon(\lambda_{p_{3}})^{\mu}*\varepsilon
(\lambda_{p_{3}})^{\nu}=-g^{\mu\nu}+\frac{p_{3}^{\mu}p_{3}^{\nu}}{m^2}
\end{eqnarray}

\begin{eqnarray}
\sum_\lambda\varepsilon(\lambda_{p_{3}})_{\perp\perp}^{\mu}*\varepsilon
(\lambda_{p_{3}})_{\perp\perp}^{\nu}=-g^{\mu\nu}+\frac{p_3^\mu
n_1^\nu+p_3^\nu n_1^\mu}{n_1\cdot p_3}-\frac{m^2n_1^\mu
n_1^\nu}{(n_1\cdot p_3)^2}
\end{eqnarray}

\begin{eqnarray}
\sum_\lambda\varepsilon(\lambda_{p_{3}})_{\|\|}^{\mu}*\varepsilon
(\lambda_{p_{3}})_{\|\|}^{\nu}=\frac{p_3^\mu
p_3^\nu}{m^2}-\frac{p_3^\nu n_1^\mu+p_3^\mu n_1^\nu}{n_1\cdot
p_3}+\frac{m^2n_1^\mu n_1^\nu}{(n_1\cdot p_3)^2}
\end{eqnarray}
Here, $n_1=(1,-\vec{p}_3)$, $n_1^2=0$, and the polarized sum rule of
$\varepsilon(\lambda_{p_{4}})$ is similar to that of
$\varepsilon(\lambda_{p_{3}})$.

\end{document}